\documentclass[12pt,preprint]{aastex}

\begin{document}

\title{A New Relation between GRB Rest-Frame Spectra and Energetics and Its Utility on Cosmology}
\author{D. Xu}
\affil{Department of Astronomy, Nanjing University, Nanjing 210093, China; xud@nju.edu.cn}


\begin{abstract}
We investigate the well-measured spectral and energetic properties of 20 gamma-ray bursts (GRBs) in
their cosmological rest frames. We find a tight relation between the isotropic-equivalent
$\gamma$-ray energy $E_{\rm iso}$, the local peak energy $E'_p$ of the $\nu F_{\nu}$ spectrum, and
the local break time $t'_b$ of the GRB afterglow light curve, which reads $E_{\rm iso}\,t'_b\propto
E'^{\,1.95\pm0.08}_p$ ($\chi^2_\nu=1.40$; $\Omega_M=0.27$, $\Omega_\Lambda=0.73$). Such a power-law
relation can be understood via the high-energy radiation processes for the GRB prompt emission
accompanying the beaming effects. We then consider this relation as an intrinsic one for the
observed GRB sample, and obtain a constraint on the mass density $\Omega _M=0.24^{+0.16}_{-0.12}$
($1\sigma$) for a flat $\Lambda$CDM universe, and a $\chi^2_{\rm dof}=1.33$ for
$\Omega_M\approx0.3$ and $\Omega_\Lambda\approx0.7$. Ongoing GRB observations in the \emph{Swift}
era are expected to confirm this relation and make its cosmological utility progress much.

\end{abstract}

\keywords{gamma rays: bursts --- cosmology: observations---cosmology: distance scale}

\section {Introduction}
Gamma-ray bursts (GRBs) are the most powerful explosions in the universe since the Big Bang. Their
cosmological origins are identified by the redshift measurements of their exploded remnants,
usually named ``GRB afterglows'', or their host galaxies. GRBs are common regarded as jetted
phenomena, supported by the observational evidence that an achromatic break appears in the
afterglow light curve, which declines more steeply than in the spherical model largely due to the
edge effect and the laterally spreading effect (Rhoads 1999; Sari et al. 1999) or the off-axis
viewing effect (Rossi, Lazzati \& Rees 2002; Zhang \& M\'esz\'aros 2002a; Berger et al. 2003). With
the advantages of huge energy release for the prompt emission and immunity to dust extinction for
the $\gamma$-ray photons, GRBs are widely believed to be detectable out to a very high redshift of
$z\sim10-20$ (Lamb \& Reichart 2000; Ciardi \& Loeb 2000; Bromm \& Loeb 2002; Gou et al. 2004).

Similar to the Phillips relation in Type Ia supernovae (SNe Ia; Phillips 1993), a tight relation in
GRBs, linking a couple of energetic and observational properties, could make GRBs new standard
candles. Amati et al. (2002) reported a power-law relation between the isotropic-equivalent
$\gamma$-ray energy $E_{\rm iso}$ of a GRB and its rest-frame $\nu F_{\nu}$ peak energy $E'_p$.
Unfortunately, the large scatter around this relation stymies its cosmological purpose. However,
after joining the burst's half-opening angle $\theta$, Ghirlanda et al. (2004a) found that a tight
relation is instead between the collimation-corrected energy $E_\gamma$ and $E'_p$, which reads
$E_\gamma\equiv E_{\rm iso}(1-\cos \theta)\propto E'^{\,\kappa}_p$ ($\kappa$ is the power).
Recently, Liang \& Zhang (2005) reported a multi-variable relation between $E_{\rm iso}$, $E'_p$,
and the local achromatic break $t'_b$ of the burst's afterglow light curve, i.e., $E_{\rm
iso}\propto E'^{\,\kappa_1}_p t'^{\,\kappa_2}_b$ ($\kappa_1$ and $\kappa_2$ are the powers).
Applying these two relations to different observed GRB samples, meaningful cosmological constraints
have been performed in a series of works (e.g. Dai, Liang \& Xu 2004; Ghirlanda et al. 2004b;
Firmani et al. 2005; Xu, Dai \& Liang 2005; Ghisellini et al. 2005; Qin et al. 2005; Mortsell \&
Sollerman 2005; Liang \& Zhang 2005; Bertolami \& Silva 2005; Lamb et al. 2005 for a GRB mission
plan to investigate the properties of dark energy).

In this paper we investigate the well-measured spectral and energetic properties of the up-to-date
20 GRBs in their cosmological frames and report a new relation between $E_X$ ($E_X\equiv E_{\rm
iso}\,t'_b$) and $E'_p$, which is $E_X\propto E'^{\,b}_p$ (where $b$ is the power). More
importantly, inspiring constraints on cosmological parameters can be achieved if this newly found
relation is indeed an intrinsic one for the observed GRB sample.

\section{Data Analysis and Statistical Result}
The ``bolometric'' isotropic-equivalent energy of a GRB is given by
\begin{equation}
E_{\rm iso}= 4\pi d_L^2 S_\gamma  k(1 + z)^{-1},
\end{equation}
where $S_\gamma$ is the fluence in the observed bandpass and $k$ is a multiplicative correction
relating the observed bandpass with a standard rest-frame bandpass (1-$10^4$ keV) (Bloom, Frail
\& Sari 2001), with its fractional uncertainty
\begin{equation}
\left( {\frac{{\sigma _{E_{\rm iso} } }}{{E_{\rm iso} }}} \right)^2  = \left( {\frac{{\sigma
_{S_\gamma } }}{{S_\gamma  }}} \right)^2  + \left( {\frac{{\sigma _k }}{k}} \right)^2.
\end{equation}
Thus, the collimation-related energy $E_X$ (see $\S 4$ of this work) is
\begin{equation}\label{ExEp}
E_X\equiv E_{\rm iso}\,t'_b=4\pi d_L^2 S_\gamma k\, t_b(1 + z)^{-2},
\end{equation}
where $t_b$ is the observed break time of the afterglow light curve in the optical band, with
its fractional uncertainty
\begin{equation}
\left( {\frac{{\sigma _{E_X } }}{{E_X }}} \right)^2  = \left( {\frac{{\sigma _{S_\gamma  }
}}{{S_\gamma  }}} \right)^2  + \left( {\frac{{\sigma _k }}{k}} \right)^2  + \left(
{\frac{{\sigma _{t_b } }}{{t_b }}} \right)^2.
\end{equation}

To calculate $k$-correction of a burst, one is required to know its spectral parameters fitted by
the Band function, i.e., the low-energy spectral index $\alpha$, the high-energy spectral index
$\beta$, and the observed peak energy $E_p\equiv E'_p/(1+z)$ (Band et al. 1993). In this paper, we
considered those GRBs with known redshifts, fluences, spectral parameters, and break times, and
thus collected a sample of 20 bursts shown in Table 1. The criterions of our data selection are:
(1) The fluence and spectral parameters of a burst are chosen from the same original literature as
possible. If different fluences are reported in that literature, we choose the measurement in the
widest energy band. If this criterion is unsatisfied, we choose the fluence measured in the widest
energy band available in other literature; (2) The observed break time for each burst is taken from
its data in the optical band; (3) When the fractional uncertainties of $E_p$, $S_\gamma$ and $k$
are not reported in the literature, they are taken to be $20\%$, $10\%$ and $5\%$, respectively;
(4) To be more reliable, a lower limit of the fractional uncertainty of $t_b$ is set to be $10\%$.

In this section, we investigate the $E_{\rm iso}-E'_p$ and $E_X-E'_p$ relations for a
Friedmann-Roberston-Walker cosmology with mass density $\Omega_M$ and vacuum energy density
$\Omega_\Lambda$. Therefore, the theoretical luminosity distance in equations (1) and (3) is given
by
\begin{eqnarray}
d_L & = & c(1+z)H_0^{-1}|\Omega_k|^{-1/2}{\rm sinn}\{|\Omega_k|^{1/2}\nonumber  \\ & & \times
\int_0^zdz[(1+z)^2(1+\Omega_Mz)-z(2+z)\Omega_\Lambda]^{-1/2}\},
\end{eqnarray}
where $\Omega_k=1-\Omega_M-\Omega_\Lambda$, $H_0 \equiv 100\,h\,\,{\rm{ km\,s^{-1}Mpc^{-1}}}$ is
the present Hubble constant, and ``sinn'' is $\sinh$ for $\Omega_k > 0$ and $\sin$ for $\Omega_k <
0$. For $\Omega_k=0$, equation (5) degenerates to be $c(1+z)H_0^{-1}$ times the integral (Carroll
et al. 1992). According to the conclusions of the \emph{Wilkinson Microwave Anisotropy Probe}
(WMAP) observations, we here choose $\Omega_M=0.27$, $\Omega_\Lambda=0.73$, and $h=0.71$ (e.g.
Spergel et al. 2003). The dashed and solid lines in Figure 1 respectively represent the best-fit
powerlaws for the $E_{\rm iso}-E'_p$ and $E_X-E'_p$ relations (weighting for the uncertainties on
both coordinates, see Press et al. 1999). We find
\begin{equation}
\frac{{E_{{\rm{iso}}} }}{{10^{52} \,{\rm{ergs}}}} = 10^{0.27\pm0.03}   \left( {\frac{{E'_p }}{{10^2
\,{\rm{KeV}}}}} \right)^{1.90\pm0.06}
\end{equation}
with a reduced $\chi^2=5.99$, and
\begin{equation}
\frac{{E_X}}{{10^{52} \,{\rm{ergs}}}} = 10^{-0.04\pm0.04}  \left( {\frac{{E'_p }}{{10^2
\,{\rm{KeV}}}}} \right)^{1.95\pm0.08}
\end{equation}
with a reduced $\chi^2=1.40$. As can be seen, although the power of the $E_{\rm iso}-E'_p$ relation
in this work is roughly consistent with that in Amati et al. (2002) derived from 9 \emph{BeppoSAX}
bursts, i.e., $E'_p\propto E^{\,0.52\pm0.06}_{\rm iso}$ ($\chi^2_\nu=0.91$), the dispersion around
this relation increases seriously. Instead, the scatter of the $E_X-E'_p$ relation is rather small.
One may notice that it is the $E_X-E'_p$ relation rather than the Amati relation whose power more
converges at $\sim2$. Based on the above statistical findings, we consider the $E_X-E'_p$ relation
for the observed GRB sample to constrain cosmological parameters.

\section{Constraints on Cosmological Parameters}
The $E_X-E'_p$ relation can be given by
\begin{equation}
E_X /10^{52} {\rm{ergs = }}10^a (E'_p /100{\rm{KeV}})^b,
\end{equation}
where the dimensionless parameters $a$ and $b$ are assumed to have no covariance. Combining
equations (3) and (8), we derive the apparent luminosity distance as
\begin{equation}
d_L  = 9.142 \times 10^{a/2} (E'_p /100)^{b/2} (1 + z)(kS_\gamma  t_b )^{ - 1/2} {\rm{Mpc}},
\end{equation}
with its fractional uncertainty being
\begin{eqnarray}
\left( {\frac{{\sigma _{d_L } }}{{d_L }}} \right)^2  & = & \left( {\frac{{\sigma _{S_\gamma  }
}}{{2S_\gamma  }}} \right)^2  + \left( {\frac{{\sigma _k }}{{2k}}} \right)^2  + \left(
{\frac{{\sigma _{t_b } }}{{2t_b }}} \right)^2  + \left( {\frac{b}{2}\frac{{\sigma _{E_p }
}}{{E_p }}} \right)^2 \nonumber \\ & & + \left( {\frac{{\ln 10}}{2}\sigma _a } \right)^2  +
\left( {\frac{{\ln (E'_p /10^2 )}}{2}\sigma _b } \right)^2,
\end{eqnarray}
where all the quantities are assumed to be independent of each other and their uncertainties
follow Gaussian distributions. The distance modulus is calculated by $\mu = 5\log (d_L
/{\rm{Mpc}}) + 25$, and its uncertainty is computed through $\sigma _\mu   = (5/\ln 10)(\sigma
_{d_L } /d_L )$.

Although the $E_X-E'_p$ relation is similar to the Phillips relation in SNe Ia, the methods by
which to constrain cosmological parameters are different (Riess et al. 1998; Firmani et al. 2005,
Xu et al. 2005). For SNe Ia, a Phillips-like relation is known to those relatively high-$z$ objects
after it has been calibrated by nearby well-observed SNe Ia, because the theoretical luminosity
distance is irrelevant with cosmological parameters, i.e. $d_L=z(c/H_0)$, when $z\ll1$. While for
GRBs, one won't know that sole set of $(a,b)$- parameters in the $E_X-E'_p$ relation until a
low-$z$ GRB sample is established\footnote{Possible cosmic evolution for this relation herein
cautioned.}. Therefore, the $\chi^2$ statistic for GRBs is
\begin{equation}
\chi ^2 (\Omega,a,b|h) = \sum\limits^{20}_{k=1} {\left[ {\frac{{\mu _{{\rm{th}}} (z_k ;\Omega|h) -
\mu _{{\rm{obs}}} (z_k ;\Omega,a,b|h)}}{{\sigma _{\mu _{{\rm{obs}}} (z_k ;\,\Omega,\,a,\,b,\,\sigma
_a /a,\,\sigma _b /b)} }}} \right]} ^2,
\end{equation}
where $\Omega\equiv(\Omega_M,\,\Omega_\Lambda)$  denotes a certain cosmology, and $h$ is taken as
0.71 in this work.

We carry out a Bayesian approach to obtain GRBs' constraints on cosmological parameters. For
clarity, we divide it into three stages with the detailed steps as follows:

\textbf{Stage I}

(1) fix $\Omega_i\equiv(\Omega_M,\Omega_\Lambda)_i$, (2) calculate $\mu_{\rm th}$ and $E_X$ for
each burst for that cosmology, (3) best fit the $E_X-E'_p$ relation to yield $(a, b)_i$ and
$(\sigma_a/a,\ \sigma_b/b)_i$, (4) apply the best-fit relation parametrized by $(a, b)_i$ and
$(\sigma_a/a,\ \sigma_b/b)_i$ to the observed sample, and thus derive $\mu_{\rm obs}$ and $\sigma
_{\mu_{\rm obs}}$ for each burst for that cosmology, (5) repeat Steps 1$-$4 from $i=1$ to $i={\rm
N}$ to obtain $\mu_{\rm th}$, $\mu_{\rm obs}$ and $\sigma _{\mu_{\rm obs}}$ for each burst for each
cosmology;

\textbf{Stage II}

(6) re-fix $\Omega_j$, (7) calculate $\chi^2(\Omega_j |\Omega _i)$ by comparing $\mu_{\rm
th}(\Omega_j)$ with $\mu_{\rm obs}(\Omega_i)$, $\sigma _{\mu_{\rm obs}}(\Omega_i)$, and then
convert it to a conditional probability, i.e., probability for $\Omega_j$ which is contributed by
the relation calibrated for $\Omega_i$, by the formula of $P(\Omega_j |\Omega _i)\propto \exp [ -
\chi^2(\Omega_j |\Omega _i) /2]$, (8) repeat Step 7 from $i=1$ to $i=\rm N$ to obtain an iterative
probability for cosmology $\Omega_j$ by $P^{\rm ite}(\Omega _j ) \propto \sum\limits_i {\exp [ -
\chi^2(\Omega_j |\Omega _i) /2]\times P^{\rm ini}(\Omega _i )}$ (here the initial probability for
each cosmology is regarded as equal, i.e., $P^{\rm ini}(\Omega)\equiv1$), (9) repeat Steps 6$-$8
from $j=1$ to $j=\rm N$ to obtain an iterative probability $P^{\rm ite}(\Omega)$ for each
cosmology;

\textbf{Stage III}

The iterative probability $P^{\rm ite}(\Omega)$ for each cosmology derived on Step 9 is no longer
equal to its initial probability $P^{\rm ini}(\Omega)$, but it has not reached the final/converged
probability $P^{\rm fin}(\Omega)$. So the following process is to (10) assign $P^{\rm ite}(\Omega)$
on Step 9 to $P^{\rm ini}(\Omega)$ on Step 8, then repeat Steps 8$-$9, and thus reach another set
of iterative probabilities for each cosmology, (11) run the above iteration cycle again and again
until the probability for each cosmology converges, i.e., $P^{\rm ite}(\Omega)\Rightarrow P^{\rm
fin}(\Omega)$ after tens of cycles.

In this method, to calculate the probability for a favored cosmology, we consider contributions of
all the possible $E_X-E'_p$ relations associated with their weights. The conditional probability
$P(\Omega_j |\Omega _i)$ denotes the contribution of some certain relation, and $P^{\rm
fin}(\Omega_i)$ weights the likelihood of this relation for its corresponding cosmology. Therefore,
this Bayesian approach can be formulized by
\begin{displaymath}
P^{\rm fin}(\Omega _j ) = {{\sum\limits_{i = 1}^{\rm{N}} {P{\rm{(}}\Omega _j {\rm{|}}\Omega _i
{\rm{)}} \times P^{\rm fin}{\rm{(}}\Omega _i {\rm{)}}} } \mathord{\left/
 {\vphantom {{\sum\limits_{i = 1}^{\rm{N}} {P{\rm{(}}\Omega _j {\rm{|}}\Omega _i {\rm{)}} \times P_{\rm GRB} {\rm{(}}\Omega _i {\rm{)}}} } {\sum\limits_{i = 1}^{\rm{N}} {P^{{\rm{fin}}} {\rm{(}}\Omega _i {\rm{)}}} }}} \right.
 \kern-\nulldelimiterspace} {\sum\limits_{i = 1}^{\rm{N}} {P^{\rm fin} {\rm{(}}\Omega _i {\rm{)}}} }} \,\,\,\,\,\,(j=1,{\rm N}).
\end{displaymath}

Shown in Figure 2 are the constraints for $(\Omega_M, \Omega_\Lambda)$ from 20 observed GRBs (solid
contours), using the likelihood procedure presented here. The GRB dataset is consistent with
$\Omega_M\approx0.3$ and $\Omega_\Lambda\approx0.7$, yielding a $\chi^2_{\rm dof}=1.33$. For a flat
universe, we measure $\Omega _M=0.24^{+0.16}_{-0.12}$ at the $68.3\%$ confidence level (C.L.). The
best-fit point is $\Omega_M=0.24$ and $\Omega_\Lambda=0.76$ (red cross). Shape of the elliptic
confidence contours trends to well constrain $\Omega_M$ and provide evidence for cosmic
acceleration with an enlarged GRB sample. For comparison, we also plot the constraints from 157
gold SNe Ia in Riess et al. (2004) (dashed contours). Closed $\Omega_M-\Omega_\Lambda$ ranges
derived from these two datasets at the same C.L.s have conveyed the advantages of high-$z$ distance
indicators in constraining cosmological parameters as well as other cosmological issues.
Additionally, GRBs are complementary to SNe Ia on the cosmological aspect. We see from the figure
that a combination of SNe and GRBs wound make the cosmic model of $\Omega_M\approx0.3$ and
$\Omega_\Lambda\approx0.7$ more favored, thus better in agreement with the result form WMAP
observations (e.g. Bennett et al. 2003). GRBs are hopeful to be new standard candles.

\section{Conclusions and Discussion}
We report a new relation between GRB rest-frame energetics and spectra, which is $E_{\rm iso}\,t'_b
/10^{52} {\rm{ergs  }}=10^{-0.04\pm0.04} (E'_p /100{\rm{KeV}})^{1.95\pm0.08}$ ($\chi^2_\nu=1.40$)
for $\Omega_M=0.27$, $\Omega_\Lambda=0.73$ and $h=0.71$. Considering this power-law relation for
the 20 observed GRBs, we find a constraint on the mass density $\Omega _M=0.24^{+0.16}_{-0.12}$
($1\sigma$) for a flat universe together with a $\chi^2_{\rm dof}=1.33$ for $\Omega_M\approx0.3$
and $\Omega_\Lambda\approx0.7$.

As previously discussed, the Ghirlanda relation reads $E_\gamma\equiv E_{\rm iso}(1-\cos
\theta)\propto E'^{\,\kappa}_p$ ($\kappa$ is the power) under the framework of uniform jet model.
Within this model, one can calculate the half-opening angle of a GRB jet by $\theta \propto
t'^{\,3/8}_b (n \eta_\gamma)^{1/8} E_{{\rm{iso}}}^{ - 1/8}$, where $n$ is the uniform circumburst
medium density, and $\eta_\gamma$ denotes the conversion efficiency of the initial ejecta's kinetic
energy to $\gamma$-ray energy release (Rhoads 1999; Sari et al. 1999). Thanks to strong collimation
for GRBs, i.e. $\theta\ll1$, the Ghirlanda relation becomes $E_\gamma \propto (E_{{\rm{iso}}}\,
t'_b)^{3/4} (n\eta _\gamma )^{1/4}$. In Ghirlanda et al. (2004a), the term of $n\eta_\gamma$ was
assumed to be highly clustered for the whole sample of 15 bursts presented there, thus the
Ghirlanda relation turns out to be $E_\gamma \propto (E_{{\rm{iso}}}\, t'_b)^{3/4}$. So if
$E_{{\rm{iso}}}\, t'_b \propto E'^{\,b}_p$ ($b\sim2$), there will be $E_\gamma \propto
E'^{\,\kappa}_p$ ($\kappa \sim1.5$). In this sense, the $E_\gamma-E'_p$ and $E_X-E'_p$ relations
are consistent with each other. However, $\eta_\gamma$ should be different from burst to burst
(e.g. $\leq1\%-90\%$), and $n$ is variable for a burst in the wind environment and may expand in
several orders (see Friedman \& Bloom 2004 and references therein). Additionally, these two
observables and their uncertainties are very difficult to be reliably estimated. Therefore, if the
Ghirlanda relation is used to measure cosmology, the resulting constraints would depend on
different input assumptions of $\eta_\gamma$ and $n$. For contrast, such difficulties have been
circumvented when applying the $E_X-E'_p$ relation. Moreover, Liang \& Zhang (2005) reported a
generalized relation between GRB rest-frame spectra and energetics could be written as
$E_{{\rm{iso}}} /10^{52}\,{\rm{ergs}} = ({0.85\pm0.21}) \times {(E'_p/{\rm
100\,KeV})}^{1.94\pm0.17} \times (t'_b /1\,{\rm{day}})^{ - 1.24 \pm 0.23}$ for a flat universe of
$\Omega_M=0.28$, using a multiple variable regression method for the 15 bursts presented there.
Their conclusion is consistent with our statistical result (see $\S2$ in this work). More loose
constraints, however, were performed when the uncertainties of the fitted parameters in this
relation were included into the error of the apparent distance modulus (see Fig 11 in Liang \&
Zhang 2005). Among the discussed three relations, the $E_X-E'_p$ relation is the simplest and the
latter two have the advantage of making the relations explicitly model-independent and eliminating
the needs to marginalize over the unknown $\eta_\gamma$ and $n$.

So what is the underlying theoretical basis for the $E_X-E'_p$ relation? At the present stage,
plausible explanations mainly include: the standard synchrotron mechanisms in relativistic shocks
(e.g. Zhang \& M\'esz\'aros 2002b; Dai \& Lu 2002), the high-energy emission form off-axis
relativistic jets (e.g. Yamazaki et al. 2004; Eichler \& Levinson 2004; Levinson \& Eichler 2005),
and the dissipative photosphere model producing a relativistic outflow (e.g. Rees \& M\'esz\'aros
2005). The scaling relation depends on the details of each model, and it resembles the
observational result under certain simplification. Although models are different, the relation they
support has made GRBs towards more and more standardized candles.

\acknowledgments XD is thankful to Z. G. Dai, E. W. Liang and B. Zhang for valuable discussions and
D. Band in explaining the observations of GRB050525a. This work is supported by the National
Natural Science Foundation of China (grants 10233010 and 10221001), and the Ministry of Science and
Technology of China (NKBRSF G19990754).

\clearpage
\begin{table}
\rotate \caption{\label{Table1}}{Sample of 20 $\gamma$-ray bursts}

\begin{tabular}{cccccccc}
\hline\hline GRB & Redshift & $E_p(\sigma_{E_p})\tablenotemark{a}$  &  $[\alpha,
\beta]\tablenotemark{a}$ &  $S_\gamma(\sigma_{S_\gamma})\tablenotemark{b}$  &   ${\rm
Bandpass}\tablenotemark{b}$  &  $t_{\rm b}(\sigma_{t_{\rm b}})\tablenotemark{c}$  &  ${\rm
References}\tablenotemark{d}$\\
   &   &  KeV  &   & $10^{-6}\, {\rm erg\, cm^{-2}}$ & KeV  &  day  &  ($E_p$,
$S_{\gamma}$, $t_{\rm b}$)\\
\hline
970828...   &  0.9578  &   297.7[59.5]    &  -0.70, -2.07   &  96.0[9.6]      & 20-2000   & 2.2(0.4)     &1,2,3\\
980703...   &  0.966   &   254.0[50.8]    &  -1.31, -2.40   &  22.6[2.26]     & 20-2000   & 3.4(0.5)     &1,2,4\\
990123...   &  1.600   &   780.8(61.9)    &  -0.89, -2.45   &  300.0(40.0)    & 40-700    & 2.04(0.46)   &5,5,6\\
990510...   &  1.619   &   161.5(16.0)    &  -1.23, -2.70   &  19.0(2.0)      & 40-700    & 1.57(0.16)   &5,5,7\\
990705...   &  0.8424  &   188.8(15.2)    &  -1.05, -2.20   &  75.0(8.0)      & 40-700    & 1.0(0.2)     &5,5,8\\
990712...   &  0.4331  &   65.0(10.5)     &  -1.88, -2.48   &  6.5(0.3)       & 40-700    & 1.6(0.2)     &5,5,9\\
991216...   &  1.020   &   317.3[63.4]    &  -1.23, -2.18   &  194.0[19.4]    & 20-2000   & 1.2(0.4)     &1,2,10\\
011211...   &  2.140   &   59.2(7.6)      &  -0.84, -2.30   &  5.0[0.5]       & 40-700    & 1.56(0.16)   &11,2,12\\
020124...   &  3.200   &   120.0(22.6)    &  -1.10, -2.30   &  6.8[0.68]      & 30-400    & 3.0(0.4)     &13,13,14\\
020405...   &  0.690   &   192.5(53.8)    &   0.00, -1.87   &  74.0(0.7)      & 15-2000   & 1.67(0.52)   &15,15,15\\
020813...   &  1.255   &   212.0(42.0)    &  -1.05, -2.30   &  102.0[10.2]    & 30-400    & 0.43(0.06)   &13,13,16\\
021004...   &  2.332   &   79.8(30.0)     &  -1.01, -2.30   &  2.55(0.60)     & 2 -400    & 4.74(0.47)   &17,17,18\\
021211...   &  1.006   &   46.8(5.5)      &  -0.805,-2.37   &  2.17(0.15)     & 30-400    & 1.4(0.5)     &19,19,20\\
030226...   &  1.986   &   97.1(20.0)     &  -0.89, -2.30   &  5.61(0.65)     & 2 -400    & 1.04(0.12)   &17,17,21\\
030328...   &  1.520   &   126.3(13.5)    &  -1.14, -2.09   &  36.95(1.40)    & 2 -400    & 0.8(0.1)     &17,17,22\\
030329...   &  0.1685  &   67.9(2.2)      &  -1.26, -2.28   &  110.0(10.0)    & 30-400    & 0.48(0.05)   &23,23,24\\
030429...   &  2.658   &   35.0(9.0)      &  -1.12, -2.30   &  0.854(0.14)    & 2 -400    & 1.77(1.0)    &17,17,25\\
041006...   &  0.7160  &   63.4[12.7]     &  -1.37, -2.30   &  19.9[1.99]     & 25-100    & 0.16(0.04)   &26,26,27\\
050408...   &  1.2357  &   19.93(4.0)     &  -1.979,-2.30   &  1.90[0.19]     & 30-400    & 0.28(0.17)   &28,28,29\\
050525a.    &  0.606   &   78.8(4.0)      &  -0.987,-8.839  &  20.1(0.50)     & 15-350    & 0.20(0.10)   &30,30,30\\
\hline
\end{tabular}

\tablecomments{(a) The spectral parameters fitted by the Band function. The fractional uncertainty
of $E_p$ is taken as $20\%$ when not reported, and the fractional uncertainty of $k$-correction is
fixed as $5\%$; (b) The fluences and their errors in the observed energy band. The fractional
uncertainty of $S_\gamma$ is taken as $10\%$ when not reported. The fluence and spectral parameters
of a burst are chosen from the same original literature as possible. If this criterion is
unsatisfied, we choose the fluence measured in the widest energy band available in other
literature; (c) Afterglow break times and their errors in the optical band. A lower limit of $10\%$
is set for the fractional uncertainties of $t_b$; (d) References in order for $E_p$ ($[\alpha,
\beta]$), $S_{\gamma}$ (${\rm Bandpass}$), and $t_{\rm b}$.}

\tablerefs{(1) Jimenez et al. 2001; (2) Bloom et al .2003; (3) Djorgovski et al. 2001; (4) Frail et
al. 2003; (5) Amati et al. 2002; (6) Kulkarni et al. 1999; (7) Stanek et al. 1999; (8) Masetti et
al. 2000; (9) Bj{\" o}rnsson et al. 2001; (10) Halpern et al. 2000;  (11) Amati 2003; (12)
Jakobsson et al. 2003; (13) Barraud et al. 2003; (14) Berger et al. 2002; (15) Price et al. 2003a;
(16) Barth et al. 2003; (17) Sakamoto et al. 2004; (18) Holland et al. 2003; (19) Crew et al. 2003;
(20) Holland et al. 2004; (21) Klose et al. 2004; (22) Andersen et al. 2003; (23) Vanderspek et al.
2004; (24) Price et al. 2003b; (25) Jakobsson et al. 2004; (26) Butler et al. 2005; (27) Stanek et
al. 2005; (28) Sakamoto et al. 2005; (29) Godet et al. 2005; (30) Blustin et al. 2005.}

\end{table}
\clearpage

\begin{figure}
\centering
\includegraphics[width=0.8\textwidth]{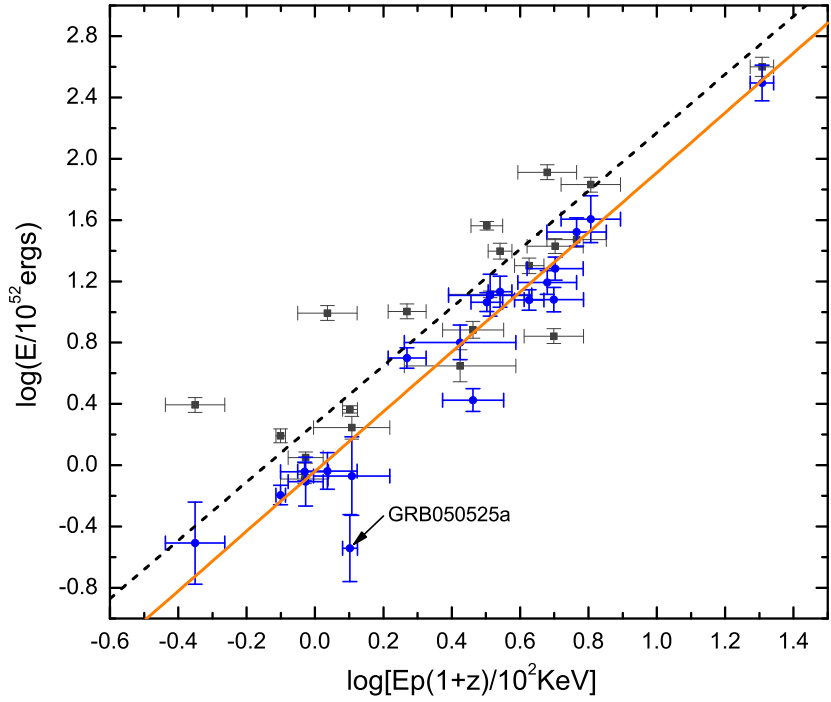}
\caption{The $E_X/E_{\rm iso}-E'_p$ plane. Circles represent $E_X$ vs. $E'_p$ for 20 GRBs, while
squares denote $E_{\rm iso}$ vs. $E'_p$. Solid and dashed lines are the best fits of the $E_X-E'_p$
and $E_{\rm iso}-E'_p$ relations, using $\Omega_M=0.27$, $\Omega_\Lambda=0.73$, and
$h=0.71$.\label{fig1}}
\end{figure}

\clearpage
\begin{figure}
\centering
\includegraphics[width=0.7\textwidth]{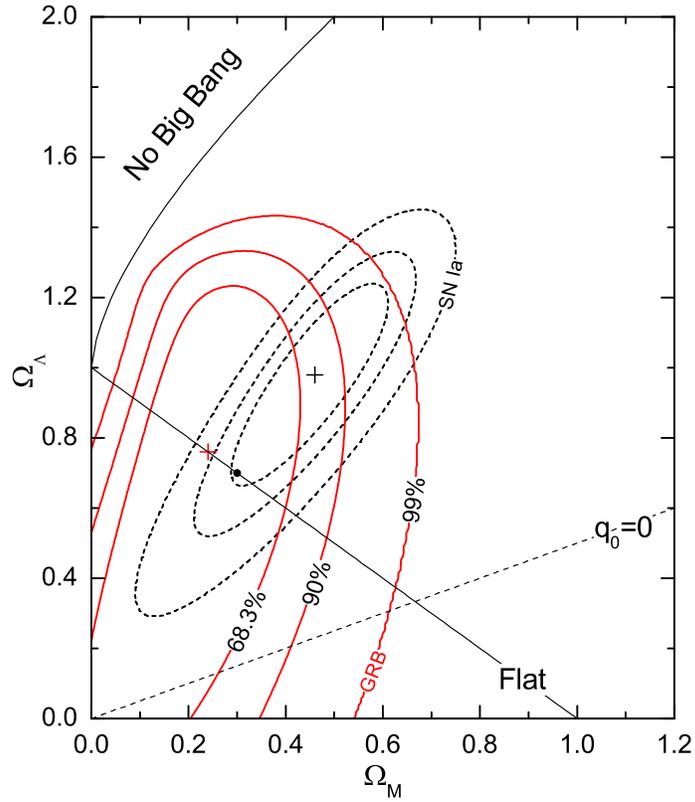}
\caption{Joint confidence intervals (68.3\%, 90\%, 99\%) in the $\Omega_M-\Omega_\Lambda$ plane
from 20 observed GRBs (solid contours), and from 157 gold SNe Ia (dashed contours). Red and black
crosses mark the best fits of the two samples, while black dot denotes the concordance model of
$\Omega_M=0.3$.\label{fig2}}
\end{figure}

\end{document}